\documentclass[12pt,a4paper,reqno]{article}
\usepackage{latexsym}
\usepackage{graphics}

\begin{document}

\newcommand{\lt}{\left}
\newcommand{\rt}{\right}
\newcommand{\nn}{\nonumber}
\newcommand{\blank}[1]{}
\newcommand{\X}{\tilde{X}}
\newcommand{\Lt}{\tilde{L}}
\newcommand{\La}{\Lambda}
\newcommand{\LdKdL}{\Lambda\frac{\partial K(p^2)}{\partial\Lambda}}
\newcommand{\LddL}[1]{\Lambda\frac{\partial #1}{\partial\Lambda}}
\newcommand{\Tr}{\mathrm{Tr}}
\begin{titlepage}
\begin{flushright}
{\bf DAMTP-2005-11} \\
hep-th/0502056 \\
\end{flushright}
\begin{center}
\vspace{.2in}
 {\large {\bf Non-Perturbative Tachyon Potential from the\\ Wilsonian Renormalization Group}}

\bigskip

\bigskip

\bigskip

%
James P. O'Dwyer\\
\vspace{.1 in}
DAMTP, Centre for Mathematical Sciences \\
University of Cambridge, Wilberforce Road \\
Cambridge CB3 0WA, England \\
{\tt j.p.odwyer@damtp.cam.ac.uk}
\vspace{.1 in}
\end{center}

\bigskip

\bigskip

\bigskip

\bigskip

\begin{abstract}

The derivative expansion of the Wilsonian renormalization
group generates additional terms in the effective $\beta$-functions
not present in the perturbative approach. Applied to the nonlinear
$\sigma$ model, to lowest order the vanishing of the $\beta$-function
for the tachyon field generates an equation analogous to that found in
open string field theory. Although the nonlinear term depends on the
cut-off function, this arbitrariness can be removed by a rescaling of
the tachyon field.

\end{abstract}

\end{titlepage}

\section{Introduction}

The calculations developed by
 Friedan and others~\cite{Honerkamp, Friedan, Alvarez} allow
 us to express world sheet scale
invariance of $\sigma$ models in terms of conditions
satisfied by target space fields. The resulting equations are
perturbative in the number of loops computed in the world sheet
field theory and are given as an expansion in powers of $\alpha'$.
For example, to lowest order the target space metric
must satisfy Einstein's
equations, while the tachyon field must solve the Klein-Gordon equation
 with negative mass-squared. In this note we apply the Wilsonian, or
Exact RG (ERG)
techniques invented by Wilson~\cite{Wilson}, Wegner
and Houghton~\cite{Wegner} and
Polchinski~\cite{Polchinski}, and developed by many others more
recently (for example see~\cite{Hasenfratz,Morris,Haagensen,Kubyshin,Bagnuls,Bervillier,Ball}
and references therein) to the same model. Earlier applications of the
 ERG to
 $\sigma$ models can be found in~\cite{Banks1, Polchinski1, Banks2,
 Brustein, Tseytlin1} and references.

\smallskip

The advantage of these ERG
techniques is that there is no reliance on a perturbation
expansion, and the
resulting conditions for scale invariance might be expected
in some sense to go beyond the known perturbative results. The
 disadvantage is that one always has to make some kind of
 approximation; the generic effective action will contain all
 field operators compatible with the symmetry of the problem, and so some
 kind of truncation is necessary to even begin a calculation.
In the next section we will write the effective action as an expansion
in world sheet derivatives, and curtail this expansion at the lowest
order. This is the local potential approximation. In terms of fields
on the target space of the sigma model, we have a tachyon field
propagating on fixed, flat geometry. Imposing scale
invariance yields equations for the tachyon field, which we will
compare to known results from open string field theory \cite{Witten1,
  Witten2,Witten3, Kutasov,
  Gerasimov}.

\smallskip

\section{The Local Potential Approximation and Tachyon Field Equation}

We regularise the theory and introduce a mass-scale $\Lambda$ 
by multiplying the propagator (with momentum $p$) by a
cut-off function, $K(\frac{p^2}{\Lambda})$, which falls rapidly to
zero for large values of the argument. The
partition function is:
\begin{eqnarray}
Z&=&\int \mathcal{D}X e^{-S[X;\La]}\nn\\
S[X;\La]&=& \int d^2\sigma\lt(
-\frac{1}{2}X^i(\sigma)K\lt(\frac{-\partial^2}{\Lambda^2}\rt)^{-1}
\partial^{2}X^i(\sigma)\rt)
+S_{int}[X;\Lambda]
\end{eqnarray}
where indices are lowered and raised using $\delta_{ij}$ and its inverse.
When we change the mass-scale $\Lambda$ the physics
is required to remain the same, and in particular the partition function
must satisfy
\begin{equation}
\Lambda \frac{d Z}{d\Lambda}=0\label{eq:inv}
\end{equation}
so that correlation functions and other physical quantities should
not depend on what scale we impose the cut-off. The additional
terms in the action needed to impose this condition are contained in
the effective interaction lagrangian, $S_{int}$. Any 
function of the fields compatible with the
symmetries of the problem may be present, and so 
in order to make progress some kind of
truncation is needed. We assume a local expansion in powers of
world sheet derivatives:
\begin{equation}
S_{int}[X;\La]=\int d^2\sigma\lt(\La^{2}T(X;\La)
+U_{ij}(X;\La)\partial_\alpha X^i
\partial^\alpha X^j +\dots \rt)\label{eq:trunc1}
\end{equation}
with powers of $\La$ inserted to make the coefficient functions dimensionless.
Further, in the Local
Potential Approximation (LPA) the world
sheet momenta are assumed to be sufficiently small so that we can neglect the
$U_{ij}$ and higher derivative terms in Equation~\ref{eq:trunc1}.

\smallskip

Writing the action in terms of Fourier-transformed world-sheet
fields $\tilde{X}(q)$, the partition function is:
\begin{equation}
Z=\int \mathcal{D}\X \exp\lt[\int d^2 q\lt(
-\frac{1}{2}\X^i(q)\X^i(-q)q^2
K\lt(\frac{q^2}{\Lambda^2}\rt)^{-1}\rt)-S_{int}[\X;\Lambda]\rt]
\end{equation}
We define dimensionless world sheet momenta $p=\frac{q}{\Lambda}$, and
from the definition of the partition function:
\begin{equation}
\Lambda\frac{d Z}{d \Lambda}= \int\mathcal{D}\X\lt(\int d^2p\
\lt[-\X^i(p)\X^i(-p)p^4 \
K(p^2)^{-2}K'(p^2)\rt]-\LddL{S_{int}}\rt)e^{-S[\X;\La]}\label{eq:LdZdL}
\end{equation}
Following Polchinski~\cite{Polchinski}, we now choose the following
condition for $S_{int}$ and demonstrate that it is sufficient to
satisfy Equation~\ref{eq:inv}:
\begin{eqnarray}
\LddL{S_{int}}=-\int d^2p
K'(p^2)\lt(\frac{\delta
    S_{int}}{\delta\X^i(p)}\frac{\delta
    S_{int}}{\delta\X^i(-p)}-\frac{\delta^2
S_{int}}{\delta\X^i(p)\delta\X^i(-p)}\rt)\label{eq:LdLdL}
\end{eqnarray}
Upon substitution of Equation~\ref{eq:LdLdL} into
Equation~\ref{eq:LdZdL} we find:
\begin{eqnarray}
\int d^2p
\ K'(p^2)\int\mathcal{D}\X \frac{\delta}{\delta\X^i(p)}
\lt[\lt[2\X^i(p)p^2K(p^2)^{-1}+
\frac{\delta}{\delta\X^i(-p)}\rt]
e^{-S[\X;\La]}\rt]\label{eq:id} =0
\end{eqnarray}
up to an overall infinite constant which can be absorbed into the
normalization of $Z$. Hence~\ref{eq:inv} is satisfied,
as claimed. The identity
above is essentially that used
in~\cite{Polchinski}, and is an application of Gauss' divergence
theorem. (As pointed out in~\cite{Polchinski}, this naive
manipulation is justified because there is a cut-off).

\smallskip

We can now translate Equation~\ref{eq:LdLdL} into conditions on
the coefficient fields in Equation~\ref{eq:trunc1}, and then
impose scale invariance. Using the LPA, we retain only $T(X)$ in
Equation~\ref{eq:trunc1}, and assume that it can be expanded as a
power series in $X$. Treating $X$ as effectively constant in
$\sigma$ (see e.g. \cite{Ball}), the derivation is then fairly lengthy but unambiguous~\cite{Golner, Ball}. We find that:
\begin{eqnarray}
0= \LddL{T} &=&\lt[-I_0\frac{\partial^2 T}{\partial
X^i\partial X^i}
  +K_0\frac{\partial
    T}{\partial X^i}\frac{\partial
    T}{\partial X^i}\rt]-2T\label{eq:erg10}
\end{eqnarray}
where
\begin{eqnarray}
K_0&=&-K'(0)\nn\\
I_0&=&
-\frac{1}{(2\pi)^2}\int d^2p\ K'(p^2)=
-\frac{1}{4\pi}\int_0^\infty du\ K'(u)
=\frac{1}{4\pi}[K(0)-K(\infty)]=\frac{1}{4\pi}
\end{eqnarray}
(For the case of a scalar field theory, when there
are no indices, these equations reduce to those derived in
e.g.\cite{Kubyshin, Ball}). Finally, writing the kinetic term
in the action in the usual way for a bosonic string,
with the conventional factor of
$2\pi\alpha'$\cite{GSW,Polchinski2}, Equation~\ref{eq:erg10} becomes:
\begin{equation}
0 =-\frac{1}{2}\alpha'\partial^2 T
  +2\pi\alpha'K_0(\partial_i
    T)^2-2T\label{eq:erg30}
\end{equation}
In a Lorentzian signature the linearized version of
Equation~\ref{eq:erg30} is just Klein-Gordon with a negative
mass-squared. The appearance of a quadratic term cannot be deduced
perturbatively, but the equation satisfied by $T$ in open string
field theory~\cite{Kutasov, Gerasimov} is identical in form. We note
that this differs from an earlier result by 
Tseytlin~\cite{Tseytlin1}, where the coefficient
$K_0$ was found to be explicitly dependent on the cut-off,
$\Lambda$. The remaining
arbitrariness of our $K_0$ is an inevitable 
consequence of the truncation of the
derivative expansion. However, so long as it is non-zero 
the arbitrariness is physically irrelevant; 
the field can always be rescaled to produce any
desired coefficient for the quadratic term. 

\smallskip

\section{Conclusions}

Having derived a tachyon field equation using the Wilsonian RG,
there remain some notable unresolved questions. 
The critical issue is whether $K_0$ is present in a more complete
ERG calculation. Also, truncating the world sheet
derivative expansion has truncated the $\alpha'$ expansion. The
latter is effectively an expansion in target space derivatives,
and it would be useful to understand fully the relation to the
world sheet derivative expansion.
It would further be interesting to understand the nature of
the connection to the boundary SFT used to analyse open string
tachyons~\cite{Witten2, Kutasov, Gerasimov, Headrick, Sen1}, and to
Tseytlin's $\sigma$ model approach to effective
actions~\cite{Tseytlin3,Tseytlin4,Tseytlin5}. 
It is likely to be necessary for consistency to consider the higher orders in
the derivative expansion, where the massless and massive fields become 
dynamical and couple to the tachyon. 
However, the privileged role of the flat metric in this
approach may mean that a covariant derivation is not possible.

\smallskip

\smallskip

\textbf{Acknowledgements:} Thanks to A.A. Tseytlin,
J. Polchinski and T. Banks for comments via email, and particularly to
H. Osborn and T.R. Morris for advice and comments on the manuscript.
This work is supported by the EPSRC.

\bibliographystyle{unsrt}
\bibliography{dSBiblio}
\nocite{*}
\end{document}